\documentclass{agujournal2019}
\usepackage{url} 
\usepackage{lineno}
\usepackage[inline]{trackchanges} 
\usepackage{soul}
\usepackage{upgreek}
\newcommand{\mum}{$\upmu$m}
\newcommand{\coo}{CO\textsubscript{2}}
\newcommand{\hoo}{H\textsubscript{2}O}
\newcommand{\Tcoo}{$T$\textsubscript{CO\textsubscript{2}}}

\graphicspath{{Figures/}}

\draftfalse

\journalname{Geophysical Research Letters}

\begin{document}

%
%


\title{Diurnal and Seasonal Mapping of Martian Ices With EMIRS}

%
%




\authors{Aur\'elien Stcherbinine\affil{1}, Christopher S. Edwards\affil{1},
    Michael D. Smith\affil{2},
    Michael J. Wolff\affil{3}, 
    Christopher Haberle\affil{1},
    Eman Al Tunaiji\affil{4},
    Nathan M. Smith\affil{1},
    Kezman Saboi\affil{1},
    Saadat Anwar\affil{5},
    Lucas Lange\affil{6},
    Philip R. Christensen\affil{5}
    }


\affiliation{1}{Department of Astronomy and Planetary Science, Northern Arizona University, 
    Flagstaff, AZ, USA}
\affiliation{2}{NASA Goddard Space Flight Center, Greenbelt, MD, USA}
\affiliation{3}{Space Science Institute, Boulder, CO, USA}
\affiliation{4}{Mohammed Bin Rashid Space Centre, Dubai, UAE}
\affiliation{5}{School of Earth and Space Exploration, Arizona State University, Tempe, AZ, USA}
\affiliation{6}{Laboratoire de Météorologie Dynamique (LMD/IPSL), Sorbonne Université, ENS, École Polytechnique, CNRS, Paris, France}





\correspondingauthor{Aur\'elien Stcherbinine}{aurelien.stcherbinine@nau.edu}




\begin{keypoints}
\item We monitor the seasonal growth and retreat of both polar caps over MY 36.
\item We monitor the presence of \coo{} ice on the surface of Mars over MY 36, through the season and through all times of day.
\item \coo\ ice appears at the surface at equatorial latitudes during the second half of the night
    around the equinoxes.
\end{keypoints}

%
%

%
%


\begin{abstract}    
Condensation and sublimation of ices at the surface of the planet is a key part of both the Martian \hoo\ and \coo\ cycles, either from a seasonal or diurnal aspect.
While most of the ice is located within the polar caps, surface frost is known to be formed during nighttime down to equatorial latitudes.
Here, we use data from the Emirates Mars Infrared Spectrometer onboard the Emirates Mars Mission to monitor the diurnal and seasonal evolution of the ices at the surface of Mars over almost one Martian year.
The unique local time coverage provided by the instrument allows us to observe the apparition of equatorial \coo\ frost in the second half of the Martian night around the equinoxes, to its sublimation at sunrise.

\end{abstract}

\section*{Plain Language Summary}   
The \hoo\ and \coo\ ices that form at the surface on Mars play an important role in the exchange between the atmosphere and the surface of the planet.
While most of the ice is located within the two polar caps that grew and shrink seasonally, ice is also known to condensate as surface frost during the night and sublimate during the day. This nighttime surface frost deposition can be observed even at equatorial latitudes.
In this paper we use data from the Emirates Mars Infrared Spectrometer onboard the Emirates Mars
Mission to detect the \hoo\ and \coo\ ices at the surface of the planet at all local times over
almost one Martian Year, which allows us to monitor both the seasonal and diurnal evolution of the
distribution of ices at the surface of Mars. We observe that nighttime \coo\ frost forms at equatorial latitudes in the second half of the night to disappear at sunrise around the Martian equinoxes.

%
%

\section{Introduction}  \label{sec:intro}
    The Martian polar caps are the main reservoirs of both \hoo\ and \coo\ ices on the surface of the red planet, at the interface between the surface and the atmosphere with an active exchange of volatiles by sublimation/condensation.
    Their seasonal growth and retreat \cite<e.g.,>{kieffer_2000, kieffer_2001, langevin_2005a, langevin_2007, appere_2011, calvin_2015, calvin_2017, oliva_2022} are an important process in both the \hoo\ and \coo\ cycles on the present-day Mars, which are two majors features of the global Martian atmospheric circulation \cite{forget_1999, montmessin_2017, titus_2017}.
    Every year, a significant fraction of the atmospheric \coo\ condensates into the seasonal polar caps which results in annual variations of about one-third of the global atmospheric mass \cite{leighton_1966, james_1992, hourdin_1993, hourdin_1995, forget_1998}.
    In addition, the presence of ice at the surface of the planet changes the albedo and the surface thermal inertia, thus affecting the energy budget at a local but also planetary scale.

    While the ices are essentially located in the polar regions, both \coo\ \& \hoo\ ices can also be observed seasonally under low latitudes, in the shadows of pole-facing slopes or at the bottom of some craters \cite{schorghofer_2006, brown_2008, carrozzo_2009, vincendon_2010, conway_2012, lange_2022}.
    From a more transient perspective, frost has also been observed to be deposed on the surface during nighttime, and remaining until early morning \cite{jones_1979, landis_2007, piqueux_2016}.
    The presence of this daily cycle of \coo\ frost has been shown to have an impact on surface processes such as gullies or slope streaks formation \cite{pilorget_2016, khuller_2021a, diniega_2021, lange_2022}.
    Thus, better constraining this diurnal frost cycle and the properties of the \coo\ ice in these regions is of importance to better understand current active processes at the surface of Mars.
    In addition, as the observations of the seasonal growth and retreat of the polar caps over the past years have shown interannual variations \cite{piqueux_2015a}, it is of interest to monitor the evolution of both the Seasonal North Polar Cap (SNPC) and Seasonal South Polar Cap (SSPC).

    In this paper, we use for the first time data from the Emirates Mars Mission to map and monitor the evolution of the Martian polar caps along with the nighttime surface \coo\ frost at low latitudes.
    First, we describe in Section~\ref{sec:methods} the data set and methods used in this study for the detection and mapping of the surface ices. Then, Section~\ref{sec:results} presents the seasonal and diurnal monitoring of the polar caps and the midlatitude nighttime \coo\ frost. Finally, Section~\ref{sec:ccl} summarizes the main points of the study.

\section{Dataset and methods}   \label{sec:methods}
    \subsection{Dataset}    \label{sec:data}
        The Emirates Mars InfraRed Spectrometer (EMIRS) instrument onboard the Emirates Mars Mission (EMM) "Hope" probe is a Fourier Transform Infrared spectrometer that is observing the Martian surface and atmosphere between 6 and 100~\mum\ with a selectable spectral resolution of 5~cm$^{-1}$ or 10~cm$^{-1}$ from February 2021 \cite{edwards_2021, amiri_2022}. The unique orbit of EMM allows EMIRS to observe the whole Martian surface across all local times in $\sim4$~orbits, which corresponds to $\sim5^\circ$ of $L_s$, or 10 Earth days.
        For every daytime EMIRS observation, high-resolution UV-visible images of the full Martian disk are acquired immediately before by the Emirates eXploration Imager (EXI) instrument \cite{jones_2021}. The coordination of these two instruments allows a direct comparison between the two datasets.

        With a pixel size typically between 100 and 300~km, it is important to consider the spatial extent of each pixel to compute accurate maps, especially for the study of the polar regions where the emission angles are high.
        However, although polygons of the pixels footprints as computed by the SPICE kernels \cite{acton_1996, acton_2018} are provided to users, their use to generate maps is not straightforward and significantly time-consuming.
        Thus, we have developed a new Python module called "SPiP" (\emph{Spacecraft Pixel footprint Projection}) that generates an approximation of the pixel footprints projected on a regular longitude/latitude map using 3D trigonometry and assuming (for now) a spherical planet \cite{stcherbinine_2023a}. 

        In this study, we use the Martian surface temperature previously retrieved using a
        multiple-step algorithm applied on a large portion of the EMIRS spectra between 7.6 and
        40~\mum\, excluding the strong CO$_2$ absorption band at 15~\mum\ \cite{smith_2022} between
        $L_s=6^\circ$ (MY 36) and $L_s=11^\circ$ (MY 37), that is, EMM orbits 8 to 323. The uncertainties on the retrieved surface temperature values have been estimated to be about 1~K \cite{smith_2022}.
        
        \subsubsection*{Data filtering}
        Then, in order to prevent our processing for instrumental bias and data artifacts, we only consider in our maps the data from pixels that meet the conditions below:
        \begin{itemize}
            \item The emission angle is lower than $80^\circ$.
            \item The retrieved surface temperature is between 140 and 300~K.
            \item The entire field of view of the pixel is within the Martian disk.
        \end{itemize}

    \subsection{Diurnally stable ice maps processing}  \label{sec:daily_maps_method}
        As ice has higher thermal inertia than the regular Martian soil (typically
        $>2{,}000$~J~K$^{-1}$~m$^{-2}$~s$^{-1/2}$ for the water ice \cite{schorghofer_2010} and
        $>1{,}000$~J~K$^{-1}$~m$^{-2}$~s$^{-1/2}$ for the \coo{} ice \cite{ciazela_2019} vs $\sim 200$~J~K$^{-1}$~m$^{-2}$~s$^{-1/2}$ \cite{putzig_2007}), we can detect the presence of surface ice that is stable across the day from the low amplitude of the surface temperature diurnal variations of these regions.
        Thus, we produce surface temperature maps from EMIRS retrievals at all Martian local times and compute the daily variations to convert them into maps of the presence of diurnally stable surface ice. 
        We converged to this temperature variations method instead of using the absolute surface temperatures in order to include \hoo\ ice in our retrievals. Although the presence of \coo\ ice can confidently be derived from the absolute surface temperature (see Section~\ref{sec:loct_maps_method}), identifying \hoo\ ice is more challenging as we have to take into account the presence of available water, and some places can exhibit temperature compatible with water ice while being actually ice-free.
        One can note that this method does not allow us to retrieve any information on the composition of the ice. Thus, for these maps, the term "ice" can refer to either \hoo\ or \coo\ ice.

        A noteworthy point: in the following, the data are duplicated on both sides of the maps (i.e., in terms of longitude) when running the interpolation and smoothing to prevent edge effects and any dependence on the choice of the center longitude.

        First, we gather all the surface temperature values retrieved from EMIRS that meet the data filtering conditions described in Section~\ref{sec:data} over 4 consecutive EMM orbits to have a full spatial coverage for all local times.
        Then, we compute surface temperature maps for bins of 1 hr of local time with a spatial resolution of $0.5^\circ \times 0.5^\circ$ by (a) projection of the pixel footprints, (b) linear interpolation of the data over the entire longitude/latitude grid with \texttt{scipy.interpolate.griddata} \cite{virtanen_2020}, and (c) Gaussian smoothing with a standard deviation $\sigma=5$.
        We also compute for each map the Gaussian kernel-density estimation (KDE) of our data for
        each map via \texttt{scipy.stats.gaussian\_kde} \cite{virtanen_2020}, and we flag as "low
        data density" all the pixels of the final maps associated with a KDE lower than $5 \times 10^{-6}$, this value has been tuned experimentally by comparing with the pixel footprints maps.

        Next, in order to increase the data coverage and prevent possible spurious temperature retrievals from some pixels, we bin the data into 3-hr maps. To do so, we crop each 1-hr map to keep only the pixels with a high data density and compute the median between the 3 consecutive maps. Then, we apply again a linear interpolation and a Gaussian smoothing filter (with $\sigma=10$) to reconstruct the data over the whole longitude/latitude grid, and we flag as "low data density" the pixels that have been reconstructed by the interpolation (i.e., without any data from one of the 3 initial cropped 1-hr maps).
        This gives us a set of 8 maps of the surface temperature for local times ranging from 00:00 to 24:00 with the following binning: 00/03/06/09/12/15/18/21/24. 

        Then, we compute a map of the amplitude of diurnal temperature variations using only the pixels for which we have at least 7 data points (over 8) across the day. We exclude the maximum and minimum values for each pixel to prevent our results from data artifacts that may affect one map and compute the difference between the maximum and minimum temperature values from the remaining data $\Delta T$. This provides us a map of the amplitude of the temperature variation between daytime and nighttime. 
        This $\Delta T$ map is then converted into a map of the ice diurnally stable at the surface from a calibration made from a comparison with EXI images.
        To do so, we compared for 9 different values of $L_s$ spanning from $L_s=58^\circ$ to $L_s=290^\circ$ the latitudinal extent of the polar caps derived from several EXI images with the corresponding $\Delta T$ maps computed with EMIRS, and identified the $\Delta T$ values in the EMIRS observations that are associated with the presence of surface ice on the EXI images acquired over the same time.
        This ice map distinguishes 3 categories:
        \begin{itemize}
            \item "Ice" for pixels associated with $\Delta T \leq25~\mathrm{K}$
            \item "Maybe ice" for pixels associated with $25~\mathrm{K} < \Delta T \leq 35~\mathrm{K}$
            \item "Not ice" for pixels associated with $\Delta T > 35~\mathrm{K}$
        \end{itemize}

        These $\Delta T$ thresholds may seem relatively high compared to what will be expected for surfaces continuously covered by ice, but due to the large size of the EMIRS pixel footprints, especially at high latitudes, some of the pixels may include a mix of icy and non-icy regions. This will tend to increase the retrieved temperature when some portion of the pixel footprints 

        Finally, we flag as "low data density" the pixels of the ice map associated with less than 7 "high data density" pixels from the 3-hr binned surface temperature maps; and we flag the entire map as "low quality" if more than 50\% of its data points are flagged as "low data density".

        These maps are also included in the L3 products released by the EXI instrument as ice masks associated with the clouds' optical depth, as the retrievals cannot distinguish between atmospheric and surface ice \cite{wolff_2022}.


    \subsection{Local time CO\textsubscript{2} ice maps}    \label{sec:loct_maps_method}
        Similarly to what is done for the diurnally stable ice maps (see section~\ref{sec:daily_maps_method}), we select all the EMIRS surface temperature values previously retrieved \cite{smith_2022} over 4 consecutive orbits, and compare them to the freezing temperature of \coo\ (\Tcoo) computed for each pixel according to Clapeyron's law \cite{piqueux_2016}:
        \begin{equation}    \label{eq:Tco2}
            \ln P = \alpha - \frac{\beta}{T\textsubscript{\coo}}
        \end{equation}
        with $\alpha=23.3494$, $\beta=3182.48$, and $P$ the \coo\ partial pressure taken as 96\% of the total surface pressure (in mbar in equation~(\ref{eq:Tco2})) \cite{piqueux_2016}.
        As P cannot be reliably retrieved from EMIRS data alone, it is taken from the Mars Climate Database (MCD) \cite{forget_1999, millour_2018} and computed to match the observational parameters of each EMIRS observation \cite{smith_2022}.
        In the following, as \coo\ is the principal component of the Martian atmosphere, the formation of \coo\ frost on the surface is not limited by the presence of gaseous \coo\ (unlike \hoo\ frost), so we will consider that \coo\ frost is present at the surface anywhere the retrieved surface temperature $T$ is lower than the predicted \Tcoo.
        However, if an EMIRS pixel is only partially covered by \coo{} ice, the retrieved temperature (averaged over the entire pixel footprint) will be higher than the expected \coo{} freezing temperature as part of the observed surface will be unfrozen and warmer than the icy part. Thus, we will only be able to detect surface \coo{} ice that entirely covers an entire EMIRS pixel.

        We project the results over the footprints of every pixel into slices of 3 hr of local time, centered on every hour of the day, with a spatial resolution of $0.5^\circ \times 0.5^\circ$.
        This results in 24 maps containing 3 values:
        \begin{itemize}
            \item "\coo\ ice" if $T \leq \Tcoo$
            \item "No \coo\ ice" if $T > \Tcoo$
            \item "Maybe \coo\ ice" if there is an overlap between pixels footprints associated with "\coo\ ice" and "No \coo\ ice" temperatures
        \end{itemize}
        
        Then, we use a nearest-pixel interpolation to reconstruct a whole map of the presence of \coo\ ice using \texttt{scipy.interpolate.griddata} \cite{virtanen_2020}.
        Finally, we compute the Gaussian KDE of our data for each map via
        \texttt{scipy.stats.gaussian\_kde} \cite{virtanen_2020}, and we flag as "low data density"
        all the pixels of the final maps associated with a KDE lower than $5 \times 10^{-6}$, this value has been tuned experimentally by comparing with the pixel footprints maps.
        One may note that consecutive maps have a temporal overlap of 1 or 2 hr, but this allows us to have a smoother view to better see when the \coo\ ice starts to form at the surface, without being biased by the choice of the local time binning.


\section{Results and discussion}    \label{sec:results}
    \subsection{Seasonal variations}    \label{sec:seasonal_variations}

        \begin{figure}[h]
            \centering
            \includegraphics[width=\textwidth]{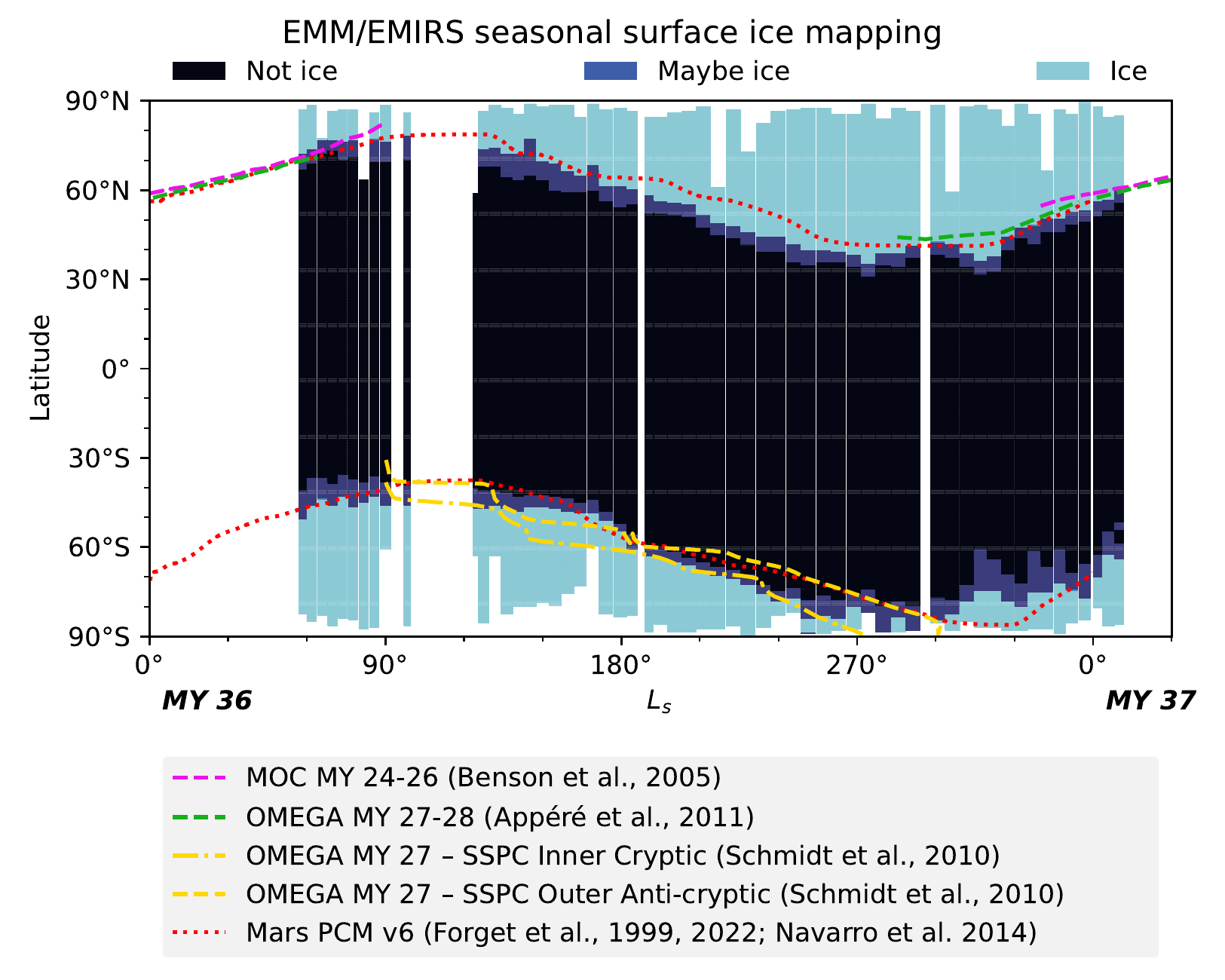}
            \caption{Seasonal variations of the latitudinal median extent of the Martian polar caps
                from EMM/EMIRS retrievals for diurnally stable surface ice. Only maps with a high-quality flag have been considered here.
                The red dotted line shows the limits of the \hoo{} polar caps derived from the Mars
                Planetary Climate Model (PCM) with a nominal dust scenario \cite{forget_1999,
                navarro_2014a}, the green and violet dashed lines show the limits of the Seasonal
                North Polar Cap derived from MOC and OMEGA observations respectively acquired during
                MY~24-26 \cite{benson_2005} and MY~27-28 \cite{appere_2011}, and the yellow dotted
                lines show the limits of the Seasonal South Polar Cap during its MY~27 recession derived from OMEGA observations \cite{schmidt_2010}.}
            \label{fig:seasonal_variations}
        \end{figure}

        Figure~\ref{fig:seasonal_variations} shows the seasonal evolution of the median latitudes of the North and South seasonal polar caps between $L_s=57^\circ$ (MY 36) and $L_s=11^\circ$ (MY 37) derived from EMIRS observations. This figure has been obtained by computing the median value of the diurnally stable ice maps (cf. Section~\ref{sec:daily_maps_method}) over all longitudes for observations with a high-quality flag. 
        The temporal range encompasses the Northern Summer and the SNPC recession/SSPC progression along with part of the Northern Winter.
        In addition, we also include in Figure~\ref{fig:seasonal_variations} the edges of the polar caps as derived from other orbital instruments (OMEGA \& MOC) for previous Martian years \cite{benson_2005, schmidt_2010, appere_2011}, and predicted by the Mars PCM version 6 numeric model assessing a "nominal dust scenario" \cite{forget_1999, navarro_2014a, naar_2021, forget_2022} as a comparison with our MY~36 EMIRS retrievals.
        For the Mars PCM, the caps areas are defined as the regions where the surface \hoo\ ice layer at LST=12 averaged over all longitudes is greater than $10^{-3}$~kg.m$^{-2}$.

        A similar figure representing the seasonal variations of the \coo{} SSPC and SNPC only
        computed from the median values of the \coo{} local time maps over the day (described in
        Section~\ref{sec:loct_maps_method}) is provided as Figure~S1 in Supporting Information S1, but will not be discussed in details here.

        \subsubsection*{North Polar Cap}
        We observe that the edge of the SNPC remains stable between $70^\circ$N and $75^\circ$N between $L_s=58^\circ$ and $L_s=143^\circ$ (i.e., during the Northern Summer, when only the perennial North Polar Residual Cap (NPRC) remains), then moves progressively equatorward to reach $\sim40^\circ$N at $L_s=250^\circ$ and remains there until $L_s=290^\circ$. Unlike most of the previous studies of the evolution of the SNPC, we do not observe here the recession of the cap but its growth, which will provide a noticeable contribution to our overall understanding of the annual cycle of the SNPC.

        Considering uncertainties of $\sim3^\circ$ in latitude due to the spatial extent of the EMIRS pixels at these latitudes our SNPC retrievals match previous measurements by MOC and OMEGA at $L_s\sim57^\circ$ (MY 36) and $L_s\sim11^\circ$ (MY 37) \cite{benson_2005, appere_2011}.
        However, we can see that after the Northern Summer solstice ($L_s=90^\circ$) a larger discrepancy occurs between our EMIRS retrievals that identify the edge of the cap around $70^\circ$N -- $75^\circ$N and the previous MOC observations that report a limit around $80^\circ$N \cite{benson_2005}.
        By looking at previously obtained maps of the NPRC we can see that it barely reaches $80^\circ$N, but we also observe the presence of an additional region of perennial water ice between latitudes $74^\circ$N and $80^\circ$N for longitudes ranging from $95^\circ$E to $245^\circ$E \cite<e.g.,>{langevin_2005a, stcherbinine_2021b}.
        Thus, as this area represents $\sim40\%$ of the longitudes below the polar cap and the EMIRS pixels can span over a few tenths of degrees in longitude under these latitudes, it is likely that the cap boundary that is detected here includes these icy deposits, which explains the mismatch with the MOC data that only consider the NPRC without the frost outlier at lower latitudes that is present between longitudes $95^\circ$E and $245^\circ$E \cite{kieffer_2001, langevin_2005a}.

        
        The comparison with the Mars PCM predictions shows an asymmetry over the year: the PCM SNPC boundary matches previous OMEGA \& MOC observations during the SNPC retreat phase ($L_s \sim 330^\circ - 80^\circ$) and EMIRS retrievals for $L_s = 330^\circ - 11^\circ$ and $L_s = 58^\circ - 74^\circ$, then it is $\sim5^\circ$ to $10^\circ$ above the EMIRS boundary during the expanding phase from $L_s=185^\circ$ to $L_s=270^\circ$.
        The PCM predicts a linear growth from $L_s=134^\circ$ to $L_s=170^\circ$ followed by a plateau until $L_s=195^\circ$ and a second expansion phase until $L_s=290^\circ$, while our results show a linear expansion from $L_s=140^\circ$ to $L_s=250^\circ$ follow by a plateau. The presence of this plateau between $L_s=170^\circ$ and $L_s=190^\circ$ in the model which is not observed in the EMIRS retrievals leads to the smaller extent of the SNPC predicted by the PCM compared to our retrievals during the second half of the expansion phase.
        One may also consider that the PCM has been run here for a "nominal dust scenario"
        \cite{forget_1999, forget_2022, millour_2022, montabone_2015}, that is, not including the specificities of MY 36. Previous studies \cite<e.g.,>{calvin_2015, piqueux_2015a} have shown the presence of interannual variations of a few degrees in latitude (typically up to 3-4$^\circ$) in the extent of the seasonal polar cap deposits for the same values of $L_s$. Thus, considering the uncertainties of $\sim3^\circ$ in latitude in our retrievals due to the spatial extent of EMIRS pixels footprints, the discrepancies observed between our retrievals and the PCM may be reflecting interannual variations of the evolution of the SNPC, with a faster expanding phase in MY 36 compared to the nominal scenario.
        Otherwise, the model matches the EMIRS and OMEGA observations during the Northern Winter ($L_s\sim290^\circ$) and the EMIRS retrievals during Northern Summer ($L_s\sim90^\circ$), thus also including the Northern icy deposits outside the primary perennial polar cap with the longitudinal average.

        \subsubsection*{South Polar Cap}
        The Mars SSPC is highly asymmetric and can be divided into two regions: the "cryptic" and the "anti-cryptic" \cite{kieffer_2000, schmidt_2010}. They do not have the same sublimation rate during the polar cap recession and do not extend to the same latitudes. In particular, during the Southern Summer, the perennial South Polar Residual Cap (SPRC) is only present in the "anti-cryptic" region \cite{langevin_2007, schmidt_2010}. Thus, considering the methodology used here to map the seasonal evolution of the polar caps with EMIRS, we expect our EMIRS latitudinal boundary for the SPRC to be located between the "cryptic" and "anti-cryptic" ones derived from OMEGA observations \cite{schmidt_2010}.

        Indeed, we observe a good agreement between the EMIRS results and the OMEGA observations during the Southern Winter ($L_s\sim 95^\circ - 130^\circ$) and the second half of the SSPC recession ($L_s \sim 200^\circ - 295^\circ$).
        Between $L_s=133^\circ$ and $L_s=190^\circ$ the EMIRS SSPC boundary is detected up to $6^\circ$ northern than the external OMEGA edge ("outer anti-cryptic"). However, the boundaries derived by OMEGA are "crocus" lines \cite{schmidt_2009, schmidt_2010}, i.e, the limits of the \coo\ deposits, while our methods capture equally all the ices that may be present on the surface (\coo\ or \hoo).
        Figure~12 from \citeA{langevin_2007} shows that between $L_s=130^\circ$ and $L_s=155^\circ$ the seasonal \hoo\ deposits extend outside the \coo\ ones, with detections up to $\sim 45^\circ$S which is consistent with our observations.

        Regarding the limits predicted by the Mars PCM, we observe that it matches our detections from $L_s=58^\circ$ to $L_s=92^\circ$ and for $L_s \sim 150^\circ - 225^\circ$. 
        Then, the PCM line is at slightly more equatorial latitudes (up to $5^\circ$) compared to our detections but matches the "outer anti-cryptic" limit derived by OMEGA \cite{schmidt_2009, schmidt_2010} from $L_s \sim 225^\circ$ to $L_s \sim 265^\circ$. Considering the uncertainties of $\sim 3^\circ$ in latitude in our retrievals, these discrepancies may be reflecting the inter-annual variations of the SSPC retreat \cite{piqueux_2015a}, especially as the PCM run does not include the specific scenario of MY 36. Or it could be the consequence of a mix between the cryptic and anti-cryptic areas of the SSPC while computing the median over all longitudes, with a predominance of the cryptic area in the EMIRS retrievals and the anti-cryptic area in the PCM results, which can be related to the higher uncertainties in term of longitude in the EMIRS retrievals at high latitudes due to the spatial extension of the pixel footprints.



    \subsection{Diurnal variations}     \label{sec:diurnal_variations}
        \begin{figure}
            \centering
            \includegraphics[width=\textwidth]{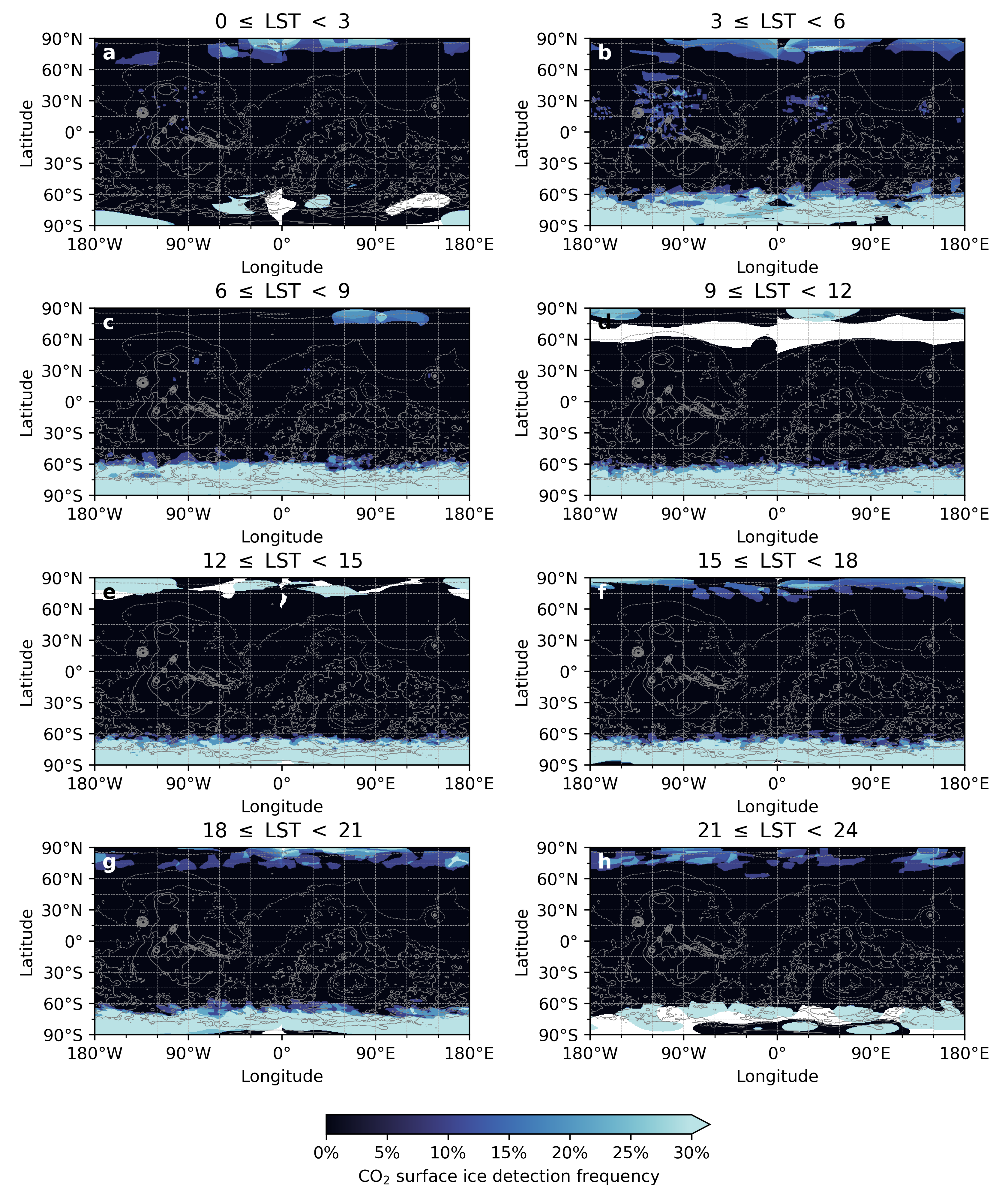}
            \caption{Frequency of detections of surface \coo\ ice from $L_s=152^\circ$ to $L_s=203^\circ$ (EMM orbits 156 to 195) for bins of 3 hr of local solar time.
            Regions without data coverage are in white.
            Pixels flagged as "low data density" are ignored here.
            We observe that \coo\ frost is detected here mostly between local solar times of 03:00 and 06:00.}
            \label{fig:diurnal_density_co2frost}
        \end{figure}

        \begin{figure}
            \centering
            \includegraphics[width=\textwidth]{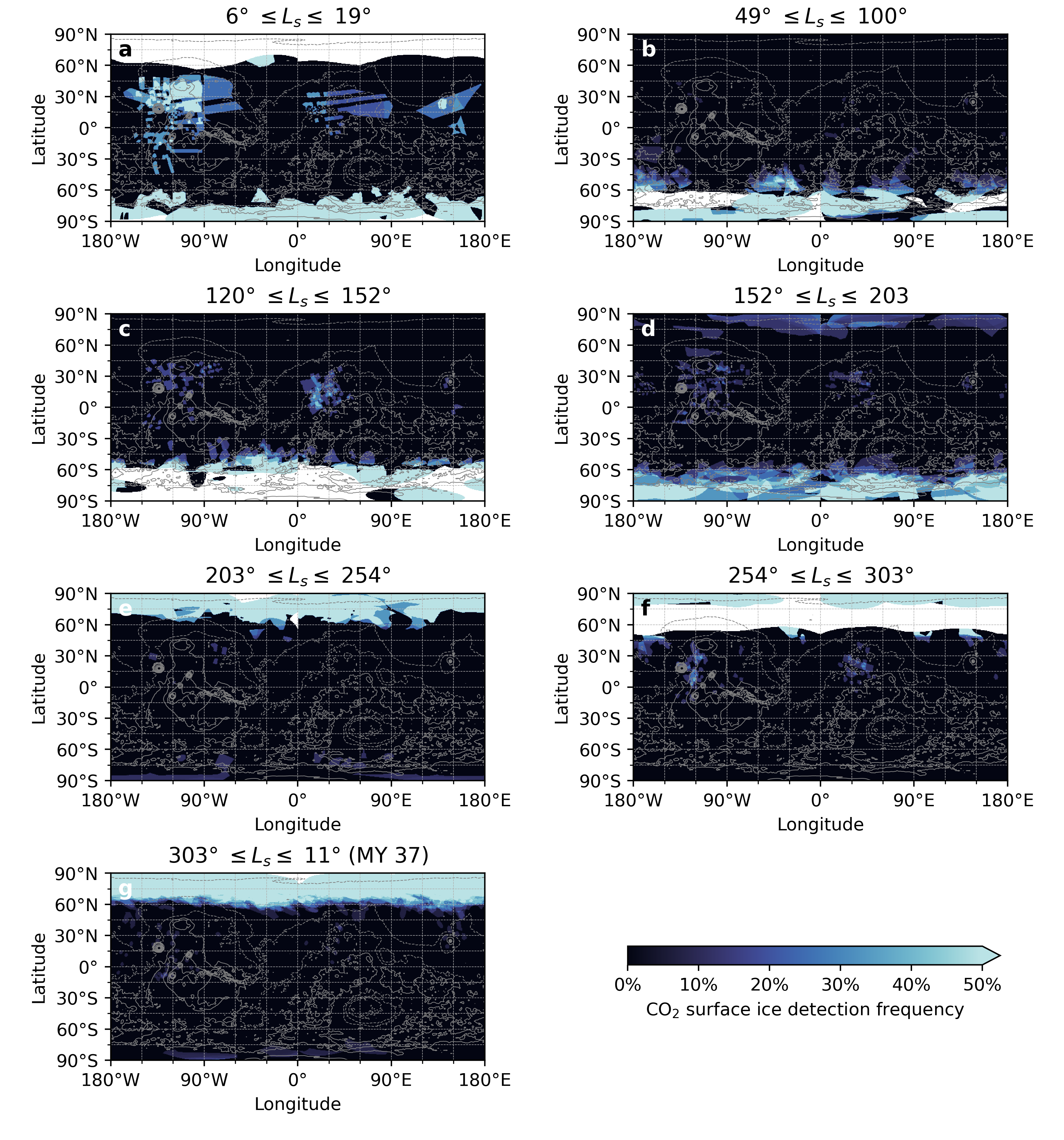}
            \caption{Frequency of detections of surface \coo\ ice between local solar times 03:00 and 06:00 for bins of $\sim 50^\circ$ of $L_s$ over MY 36.
            Regions without data coverage are in white.
            Pixels flagged as "low data density" are ignored here.}
            \label{fig:seasonal_density_co2frost}
        \end{figure}

        From the surface \coo\ ice maps generated for bins of 3 hr of LST on a 4-orbits basis as described in Section~\ref{sec:loct_maps_method}, we compute for each bin of local time a map of the percentage of \coo\ frost detections over wider ranges of $L_s$, considering only the pixels flagged as "high data density".

        Figure~\ref{fig:diurnal_density_co2frost} shows the frequency of \coo\ surface ice detections for each local time between $L_s=152^\circ$ and $L_s=203^\circ$. We can see that apart from the polar caps, \coo\ ice is detected at night under equatorial and mid-latitude regions mostly around $100^\circ$W (Tharsis region), and also around $40^\circ$E (Arabia Terra), which corresponds to low thermal inertia areas \cite{putzig_2007} where nighttime surface \coo\ frost have been detected through MCS or THEMIS measurements \cite{piqueux_2016, khuller_2021a, lange_2022}.
        Regarding the daily evolution of these \coo\ frost deposits, we observe that it starts to
        appear at midnight (panel a) and last after 6~a.m. to disappear during daytime, with a maximum of intensity between LST of 03:00 and 06:00 (panel b). 
        Condensation of the \coo\ occurs during the second half of the Martian night until sunrise.

        Figure~\ref{fig:seasonal_density_co2frost} shows the frequency of detections of surface
        \coo\ ice between 3 and 6~a.m. by EMIRS from $L_s=6^\circ$ to $L_s=290^\circ$ for 6 temporal bins of $\sim 50^\circ$ of $L_s$.
        One may note that panel a of Figure~\ref{fig:seasonal_density_co2frost} exhibits some large and awkwardly shaped detection spots of \coo{} ice, which are due to interpolation effects on maps generated with a lower data coverage at the beginning of the mission.
        We can see that our detections of non-polar surface \coo\ frost vary with the $L_s$: while they remain located in the same two regions mentioned above, the frequency of detections along with the size of the area covered by the frost evolve.
        Indeed, equatorial and mid-latitudes \coo\ frost is mostly detected for $6^\circ \leq L_s \leq 19^\circ$ and $120^\circ \leq L_s \leq 203^\circ$ (panels a, c \& d).
        Panel b ($49^\circ \leq L_s \leq 100^\circ$) corresponds to the end of Spring and early
        Summer in the Northern hemisphere, that is, the period of the year with the highest temperatures in the Northern hemisphere. Thus, as the regions where nighttime \coo\ frost is usually detected under midlatitudes are located in the Northern hemisphere, there are almost no detections outside the South polar cap during this period. Even the retrieved nighttime temperatures are not low enough to allow the formation of \coo\ ice at the surface of the planet in these regions.
        This formation of \coo\ frost late during the night with a quick sublimation at sunrise was expected from the models \cite<e.g.,>{lange_2022} with implications on the formation of gullies and slope streaks \cite{pilorget_2016, khuller_2021a, lange_2022}. But as previous nighttime observations were conducted only around 03:00, this is the first time we are able to confirm it with direct observations of the diurnal cycle of the surface \coo\ frost at low latitudes.

        Similarly, we also report almost no midlatitudes \coo\ frost detections between $L_s=203^\circ$ and $L_s=290^\circ$ (panels e \& f). This period corresponds to the Southern spring and summer, but also to the part of the Martian Year where the planet is closer to the Sun as the perihelion is reached at $L_s=251^\circ$. Thus, because of the eccentricity of the Martian orbit, the global temperature at the surface of the planet is significantly higher during the second half of the year \cite<e.g.,>{smith_2006a,bell_2008}, which explains the fewer nighttime \coo\ frost detections at equatorial and midlatitudes.

        This seasonal trend was previously observed by MCS and THEMIS \cite{piqueux_2016, khuller_2021a}. However, one may note the fewer frost detections in our EMIRS results, which may be related to the difference in terms of spatial resolution between the instruments.
        Indeed, the spatial resolution of the EMIRS pixels is comprised between 100 and 300~km \cite{edwards_2021}, which is much larger than the few km of the MCS footprints \cite{mccleese_2007} or the 100~m of THEMIS \cite{christensen_2004}.
        Plus, the detections made by both MCS and THEMIS reveal that the \coo\ frost detections may 
        be localized to areas that are sub-pixel at the EMIRS resolution \cite{piqueux_2016, khuller_2021a}. Thus, if an EMIRS pixel is only partially covered by surface \coo\ frost, the retrieved temperature, averaged over the entire footprint, will be higher than the \coo\ freezing temperature as a portion of the footprint will be unfrozen.
        This results in a lower sensibility in the surface \coo\ frost retrieving process, but also strengthen our detections as they testify of the presence of \coo\ ice over all our pixel footprints.

        


\section{Conclusion}    \label{sec:ccl}

    In this paper, we present the results of our study on the monitoring of the Martian surface ices with the EMIRS instrument onboard EMM. From the variation of the surface temperature over the day, we developed a method to automatically map the presence of diurnally stable ice at the surface, which allows us to monitor the seasonal variations of the two polar caps. Then, based on the method previously developed for MCS data \cite{piqueux_2016}, we use the surface temperature to detect and map the presence of \coo\ frost at the surface of the planet. 
    We confirm the diurnal apparition and disappearance of surface \coo{} frost under equatorial latitudes previously reported by \citeA{piqueux_2016} and observationally monitor for the first time its evolution as a function of the local time thanks to the unique orbit of the EMM probe.

    We monitor the evolution of the seasonal polar caps from $L_s=57^\circ$ (MY 36) to $L_s=11^\circ$ (MY 37), with a temporal resolution of $5^\circ$ of $L_s$ (10 Earth days).
    Plus, the large-scale view of EMIRS and the automatization of our method allow us to continuously and simultaneously monitor the annual variations of both polar caps.

    Also, we are able to observe for the first time the apparition and disappearance of low-latitude
    nighttime \coo\ frost at the surface of the planet, thanks to the unique ability of EMM
    instruments to provide full coverage in terms of local time. \coo\ ice is detected at the
    surface down to the equator around spring and fall equinoxes in the second half of the night
    (essentially between 3 and 6~a.m.) with a quick sublimation at sunrise, which confirms previous model expectations.


\section*{Data Availability Statement}
The SPiP module is freely available on GitHub at \url{https://github.com/NAU-PIXEL/spip} \cite{stcherbinine_2023a}.

Data from the Emirates Mars Mission (EMM) are freely and publicly available on the EMM Science Data Center (SDC, \url{http://sdc.emiratesmarsmission.ae}).
This location is designated as the primary repository for all data products produced by the EMM team and is designated as long-term repository as required by the UAE Space Agency.
The data available (\url{http://sdc.emiratesmarsmission.ae/data}) include ancillary spacecraft data, instrument telemetry, Level 1 (raw instrument data) to Level 3 (derived science products), quicklook products, and data users guides (\url{https://sdc.emiratesmarsmission.ae/documentation}) to assist in the analysis of the data.
Following the creation of a free login, all EMM data are searchable via parameters such as product file name, solar longitude, acquisition time, sub-spacecraft latitude \& longitude, instrument, data product level, etc.

Data products can be browsed within the SDC via a standardized file system structure that follows the convention:
\texttt{/emm/data/\textless Instrument\textgreater /\textless DataLevel\textgreater /\textless Mode\textgreater /\textless Year\textgreater /\textless Month\textgreater }

Data product filenames follow a standard convention:\linebreak
\texttt{emm\_\textless Instrument\textgreater \_\textless DataLevel\textgreater \textless StartTimeUTC\textgreater \_\textless OrbitNumber\textgreater \_\textless Mode\textgreater \_\textless Description\textgreater\linebreak
\_\textless Kernel-Level\textgreater \_\textless Version\textgreater .\textless FileType\textgreater }

EMIRS data and users guides are available at: \url{https://sdc.emiratesmarsmission.ae/data/emirs}

The Mars PCM v6 and the MCD are available from \url{http://www-mars.lmd.jussieu.fr}

\acknowledgments
The authors want to thank Sylvain Piqueux (JPL) for his helpful discussion regarding the \coo\ ice retrievals.

This work was funded by the Emirates Mars Mission project under the Emirates Mars Infrared
Spectrometer instrument via The United Arab Emirates Space Agency (UAESA) and the Mohammed bin
Rashid Space Centre.

\bibliography{biblio_stcherbinine2023_emirs_ice}

\end{document}